# Comparison of accuracy and efficiency of three-dimensional discrete ordinates and voxel-scoring Monte Carlo dose calculations for an $^{125}$I seed

**George M. Daskalov[&], R.S.Baker[@], D.W.O. Rogers[+], and J.F. Williamson[*]**

[&]Ottawa Regional Cancer Centre,
503 Smyth Rd, Ottawa, ON K1H 1C4, CANADA

[+]National Research Council of Canada, IRS/INMS,
1500 Montreal Rd., Ottawa, ON K1A 0R6, CANADA

[@]Los Alamos National Laboratory, MS D409,
Los Alamos, NM 87545, USA

[*]Department of Radiation Oncology, Virginia Commonwealth University
401 College Street, B-129, PO Box 980058, Richmond, VA 23298, U.S.A.

Corresponding author:
Jeffrey F Williamson, Ph.D.
Radiation Oncology Department
Virginia Commonwealth University
Richmond, VA 23298, U.S.A.

E-mail: jfwilliamson@vcu.edu

**ABSTRACT:**

This study compares the efficiency of three-dimensional deterministic discrete-ordinates calculations and voxel-based Monte Carlo simulations of the dose distribution around a model 6702 $^{125}$I seed in a homogeneous water phantom. The computational efficiency of the PARTISN multigroup discrete ordinates neutral particle transport code is compared to the most efficient Monte Carlo voxel-scoring computer code available to us at the time, which was developed specifically for brachytherapy. The difficulties of such comparisons from the fact that the Monte Carlo and discrete ordinates method exhibit stochastic and systematic errors of different origins. To address this problem, we propose a parametric model for separating the systematic and stochastic error components of each method. Based on this model, a procedure for assessing efficiency of the two methods is proposed consisting of the following steps: (i)-apply the developed error model to evaluate the accuracy of each method relative to unbiased and accurate Monte Carlo point estimator calculations in the same phantom geometry; (ii)-develop systematic and stochastic error criteria which specifies equivalent accuracy of the two methods; and (iii) determine the relative efficiency of two methods by taking the ratio of CPU times required to satisfy these accuracy criteria. This process demostrates that three-dimensional PARTISN discrete ordinate simulations is a factor-of-two more efficient than our voxel-based Monte Carlo code in the $^{125}$I energy range. This suggests that discrete ordinates simulations can support efficiency competitive with that of Monte Carlo in three-dimensional geometries and can serve as an efficient and accurate dose-calculation engine in treatment planning for brachytherapy and other treatment modalities utilizing low energy photon fields.

## I. INTRODUCTION.

The Discrete Ordinates Method (DOM) is a deterministic solution of the general Boltzmann equation governing the particle transport [1,2]. DOM systematically discretizes the energy, angular, and spatial domains of the problem phase space and the associated cross-section data. Our work[1,3-5] was the first to evaluate the accuracy and feasibility of multidimensional DOM simulations for brachytherapy-dosimetry problems. Our two-dimensional simulations demonstrated that the DANTSYS multigroup-DOM neutral-particle transport code supported dose computation accuracy, relative to Monte Carlo benchmarks, oof 1%-2% over the entire range of photon energies applicable to brachytherapy[1,3-4]. We also demonstrated that the DANTSYS accuracy is maintained when its computational parameters are optimized for computational efficiency through the development of a three-group cross section library for $^{125}$I dosimetry and treatment planning.[5] The three-group DANTSYS calculations were 100-fold faster than EGS4 Monte Carlo simulations for two-dimensional problems[5], indicating the potential of the DOM for providing efficient and accurate dose calculations for low-energy brachytherapy treatment planning. However, the relative efficiency of deterministic transport methods as compared to Monte Carlo simulations is problem and code-implementation dependent. In general, the deterministic transport calculations usually offer significant speed advantages over Monte Carlo in one and two spatial dimensions, in three-dimensional calculations where information is required everywhere, and in time-dependent calculations where information is required about the time derivative[6-9]. However, these comparisons are very problem dependent[6], and, to our knowledge, no such comparisons have been applied to brachytherapy calculations.

The question we are seeking answer is: "Is DOM computational efficiency competitive with the most efficient Monte Carlo (MC) photon-transport code available for simulation of brachytherapy three-dimensional dose distributions?" Probably no general answer exists, since relative efficiency depends significantly on many code-, algorithm-, and computer-implementation variables, including variance-reduction tools deployed, degree of parallelization, and type of adaptive meshing deployed. Few rigorous quantitative studies of relative DOM and MC efficiency are available despite extensive use of DOM for deep penetration shielding calculations and other applications[1, 2, 6-9]. The main contributions of this paper are (1) a process for determining the distribution of systematic and random errors of DOM and MC over a 3D dose grid; (b) quantitative accuracy requirements for determining relative efficiency of two different transport codes; and (c) to compare efficiencies of state-of-the-art (ca. 2001) DOM and voxel-scoring Monte Carlo tools for computation of three-dimensional dose distributions around a model 6702 $^{125}$I seed.

In this work, the accuracy of DOM or the voxel-based Monte Carlo is assessed relative to corresponding Monte Carlo simulations using unbiased point-dose estimators with small statistical uncertainties. We develop a comprehensive model for systematic and statistical error deconvolution and apply it for quantitative analysis of the accuracy achieved by each method. We develop comparative accuracy criteria and apply them uniformly to both methods. DOM and voxel Monte Carlo simulations are run until these criteria are satisfied by both algorithms. Finally, we compare the computational efforts for achieving this accuracy by the mean of comparing the CPU times needed and make conclusions for the relative efficiency of each method.

## II. MATERIALS AND METHODS

### II.A. Overall methodology.

Comparing DOM and voxel-scoring Monte Carlo simulations with respect to accuracy and efficiency is complicated by the fact that the two transport solutions are susceptible to errors and artifacts of different origins. Systematic errors unique to DOM arise from the finite-difference representation used to approximate the underlying phase space and to compute the associated photon fluxes and their derivatives. Using this representation, the photon-transport equation is replaced by a set of algebraic equations that are solved iteratively[1,2]. These artifacts differ significantly from Monte Carlo voxel-scoring artifacts arising from the averaging the dose distribution over the spatial region contained within each voxel. On the other hand, because of the stochastic nature of the Monte Carlo simulation process, all Monte Carlo dose estimates are subject to statistical errors, which have a complex dependence on voxel size, photon fluence, and the estimation and variance-reduction methods employed. DOM simulations are also subject to statistical errors arising from use of stochastic ray-tracing first-collision sources used to mitigate ray effects[10]. Ray effects, or undulations in the 2D and 3D angular flux distributions due to the finite number of angular directions in the phase-space mesh, are especially prominent for discrete primary-photon sources. Replacing the discrete primary-photon source by a spatially broad and continuous first-collision source, implemented via standard Monte Carlo methods, is a more efficient alternative to using very fine directional grids[30]. At any point in the simulation domain, a specified error criterion can be fulfilled by refining the grid dimensions to reduce the systematic error and increasing the number of particle histories to reduce statistical error. Since both DOM and MC artifact magnitudes are location-dependent, the CPU-time ratios needed to achieve comparable overall accuracy will be sensitive to the region selected.

To overcome these difficulties, it is necessary to define a more general process for quantifying the spatial distribution of both the systematic and stochastic errors without assuming they have the same distribution for the two types of simulations. We are pursuing the following approach:

- Development of a generalized model for systematic and stochastic error deconvolution applicable to both DOM and the voxel-scoring Monte Carlo algorithms where error is determined at each calculation point relative to unbiased Monte Carlo point-estimator results in the same phantom geometry.
- Using error distributions measured derived from the model above, develop systematic and stochastic error criteria that can be uniformly applied to each simulation in order to establish that the same accuracy has been achieved by both DOM and MC. The resulting systematic error criterion, as well as the root mean square (RMS) error criterion used in the analysis, and their application are discussed in section II.E.

With the developed error modeling tools in place, the efficiency comparison becomes a three-step program:

1. First, Monte Carlo simulations are performed using Bounded Next Flight (BNF) point-estimators[11] to evaluate the dose distribution in water medium around the model 6702 $^{125}$I seed in 3D Cartesian geometry. The resultant dose distributions serve as accuracy benchmarks for both DOM and voxel-scoring Monte Carlo since the BNF results have been shown to be unbiased estimates of the dose at a point (within 1%)r[11]. The details of the BNF Monte Carlo calculations are discussed in subsection II.B.1.

2. Both DOM and the voxel-scoring Monte Carlo simulations are made to estimate the dose distributions in the same water phantom. The obtained results are compared to the BNF benchmarks to evaluate the spatial distributions of errors, relative to BNF MC, arising with each method. The details of voxel-scoring Monte Carlo and DOM are discussed in subsections II.C and II.D, respectively.
3. The accuracy and efficiency comparisons follow these simple rules: **(i)** Large numbers of histories (Voxel-scoring Monte Carlo) and first collision source rays (DOM) are simulated so that stochastic errors are very small. The Monte Carlo voxel sizes and the finite-differencing phase-space representation for DOM are adjusted so that the two simulations achieve the same distribution of residual systematic errors (see subsection IID). This is achieved by trial-and-error with consecutive refinements and coarsenings of the voxel sizes (Monte Carlo) or the phase-space representation. **(ii)** Then, the CPU time is adjusted by reducing the number of histories, or first collision source rays, to the minimal number which would satisfy the overall root-mean square statistical error criterion also discussed in subsection II.E; and **(iii)** finally, the CPU time obtained when both criteria are satisfied with each method is assumed to be the CPU time required for either DOM or the voxel-scoring Monte Carlo to achieve the necessary accuracy. Comparing the CPU times on the same machine is used to evaluate the relative efficiency of both methods.

### II.B. Monte Carlo Simulations.

The bounded next-flight as well as the voxel-based Monte Carlo estimators are available within our in-house Monte Carlo code, PTRAN, Version 7.6 [12-18] and can be invoked by setting the appropriate code control parameter. The Monte Carlo code was applied to the Model 6702 $^{125}$I seed in the center of a rectangular homogeneous water phantom of dimensions 20 cm x 20 cm x 20 cm. The seed geometry model used is the same one used in our previous work[5] and is shown in figure 1a. The macroscopic cross sections used in the simulations, as well as the normalized probabilities for coherent scattering, incoherent scattering, and photoeffect interactions, were derived from the DLC-146 cross-section library[19]. Characteristic X-ray production in all Monte Carlo simulations was suppressed since it is not modeled by our multigroup DOM calculations. The mass energy absorption coefficients of Hubbell and Seltzer[20] were used to convert the photon energy flux into collision kerma. Due to negligible attenuation of the $^{125}$I photons over the range of secondary electrons set in motion, secondary charged particle equilibrium can be assumed[5,21] so that absorbed dose can be accurately approximated by collision kerma[5,13,21]. Because of this equivalence we are using absorbed dose, collision kerma, and kerma interchangeably throughout this study.

#### *II.B.1. Calculations with the Bounded Next Flight point estimators (MCBNF).*

Our bounded next-flight point-estimator simulations are referred to below as MCBNF. All quantities associated with them will have the superscript MCBNF for clarity. We used this estimator to provide a benchmark dose distribution that is relatively free of both systematic and statistical errors. It does not average photon trajectories over a finite volume detector, but seeks to estimate kerma-at-a-point. Previous work[11] identified operating parameters that maximize its computational efficiency while limiting its systematic error to less than 1%.

To achieve standard errors of the mean (67% confidence interval) ranging from 0.2% to 0.6% depending on the position of the estimator, 4,000,000 and 12,500,000 histories were simulated. The MCBNF point detectors estimators were grouped in 6 sets, each consisting of 6335 detectors distributed over a spherical surface at constant distances of 0.25 cm, 0.50 cm, 1.0 cm, 2 cm, 3 cm, and 5 cm from the seed center. As shown in figure 1b, the 6335 BNF detectors at each distance were distributed uniformly over 35 longitudinal great circles (181 each) so as to uniformly span the range of polar angle, $\theta$, from 0° to 180°. The MCBNF dose rates had units of cGy/(mCi·h) and was used in this form to evaluate DOM accuracy (1 $cGy\cdot mCi^{-1}\cdot h^{-1}=7.508\times10^{-14}\ Gy\cdot Bq^{-1}\cdot s^{-1}$). The $^{125}I$ photon source spectrum used is the one given by NCRP[22].

***II.B.2. Voxel-scoring Monte Carlo simulations (3DMC).***

Voxel-scoring Monte Carlo simulations are referred to below as 3DMC. 3DMC simulations use the same geometric model and photon-interaction physics as the MCBNF simulations discussed in subsection II.B. The only difference was scoring, which uses the expected value track-length estimator, described in detail in the Appendix of reference 12, to a uniform 3D voxel grid. Our implementation uses the fast parametric ray-tracing algorithm due to Siddon[23]. Our implementation requires a uniform grid plane distance along any coordinate axis' direction. The ray tracing through the grid takes between 50% to 80% of the overall CPU time, depending upon the grid size, the number of grid planes (N) per dimension, and the total number of voxels ($N^3$), but is well suited to this application as CPU time scales with $N \log N$. To overcome the fixed grid size limitation, two 3DMC simulations were performed: one with a voxel size of 0.1 mm for distances smaller than 1 cm, and another with a voxel size of 1 mm for all other distances. These voxel sizes were chosen as the largest sizes yielding acceptable systematic error (subsection II.D) at the smallest distances of the corresponding regions (0.25 cm for 0.1 mm voxel size, and 1 cm for 1 mm voxels) and these voxel grids are referred to below as the basic voxel grids. The 3DMC scoring results are recorded in a binary file containing each voxel position and its score in units of cGy/(mCi·h). The major reason for using two different voxel grids to cover the entire phantom volume is that the number of voxels needed to span the entire $20\times20\times20 cm^3$ with uniformly small 0.1 mm voxels would exceed significantly the memory (1 GB) of our largest workstation. Therefore, we limited $N^3$ in each 3DMC simulation to 7,880,599 (3D grids of 199x199x199 voxels) thus reducing the memory requirements to approximately 850 MB. The resulting dose rates $\dot{D}^{3DMC}(x,y,z)$ at each BNF position $(x,y,z)$ were obtained by 3D linear interpolation of the grid values after factoring out the geometry factor (inverse-square law) as described in our previous work[5]. The corresponding statistical uncertainty of the interpolant was obtained in a similar manner from the uncertainties of the interpolated 3D grid results.

We analyzed 9,000,000 histories with the 0.1 mm size grid and 3,000,000 histories with the 1 mm grid in order to obtain standard deviations of the mean $\frac{\sigma_i^{3DMC}}{\dot{D}_i^{3DMC}}$ for each voxel $i$ of 0.01 or better (referred to below as high statistics) and determine the systematic error with sufficient accuracy.

**II.C. PARTISN DOM calculations.**

We distinguish the results of our DOM simulations below by assigning the superscript DOM to all quantities associated with them. The DOM computer code system PARTISN[24-26] , version 1.38, was implemented in our study. PARTISN was developed from the DANTSYS code family (used in our previous studies). It incorporates higher-order spatial-differencing schemes, adaptive mesh refinement (AMR), and supports parallel multiprocessor implementation (none of these features were used in this study). We used a three-energy bin multigroup cross-section library[5] that had been previously optimized for efficient $^{125}$I seed dose calculations. The three-group library is derived from the same DLC-146 cross section library as used by MCBNF and MC3D. Group flux-to-dose conversion factors, also derived from DLC-146, were used to convert the DOM flux into dose rates. The details of the DOM approach for radiation transport solutions, together with the typical operational parameters, have been reviewed earlier[1,3-5]. Our 3-group energy representation includes all $^{125}$I primary photons in the first group[5]. For fair comparison with the 3DMC simulations, two PARTISN simulations were performed: one to span distances less than or equal to 1 cm, and another one for distances greater than 1 cm. As discussed in II.E, different grid sizes were used for distances of 0.1 mm to 5 mm and 0.1 mm to 5 cm. However, both grids used the same spacing in the immediate vicinity of the source, so that both simulations used the same 3D spatial discretization for seed modeling shown in figure 2.

Because of the symmetry of the problem, we have carried out the PARTISN simulations in 1 octant of the 3D space and employed reflective boundary conditions along the coordinate axes as shown in figure 2. The corresponding CPU times were multiplied by a factor of 8 to account for the fact that 3DMC populates all 8 octants with histories. We employed a third order Legendre polynomial expansion ($P_3$) for the flux in the primary photon energy group – group 1, while second order polynomial expansion ($P_2$) was used for the scattering anisotropy in energy group 2. Calculations in the last energy group 3 were suppressed since the contribution of the scattered flux in this group is negligible[5]. The seed geometry and material composition were mapped into the rectangular spatial mesh representation by the auxiliary code FRAC-in-The-Box[27]. We used 36 angles ($S_{16}$) for angular discretization during the iterations within energy group 1 while 21 angles ($S_{12}$) were used for group 2. In the same manner as in our previous work[1,3-5], we used the adaptive weighted diamond-differencing (AWDD) method for space/angle discretization[28] and employed the diffusion-synthetic acceleration (DSA) merthod[29] to accelerate the convergence of the iterative source solution.

Table 1 shows the propagation of the statistical error from the primary to total kerma. As expected, the magnitude of the standard deviation of the total and primary kerma is the same in each case since the DOM scattered-photon dose component is deterministically calculated. Consecutively, the relative standard deviation of the total kerma is the primary kerma modified by the primary-to-total kerma ratio. Based on this result we estimate the statistical error of each DOM dose rate by multiplying the corresponding first-collision source statistical uncertainty by the primary-to-total kerma ratio in each spatial mesh cell. The DOM dose rate $\dot{D}_{k,i}^{DOM}(x,y,z)$ at the location of $i^{th}$ BNF detector in $k^{th}$ detector ring were obtained by the same nonlinear 3D linear interpolation procedure as for interpolating 3DMC dose grid onto the MCBNF grid.

**II.D. The error model and accuracy criteria.**

An important part of this study is to analyze and quantitatively estimate the accuracy of each simulation. Following Kawrakow and Fippel[32], we consider the quantity

$$q_i = \frac{\dot{D}_i^{MCBNF}(x,y,z) - \dot{D}_i^{Model}(x,y,z)}{\sigma_i(x,y,z)}, \tag{1}$$

where $\dot{D}_i^{MCBNF}(x,y,z)$ is the water dose rate estimated by the i-th BNF point estimator with Cartesian coordinates *(x,y,z)*, $\dot{D}_i^{Model}(x,y,z)$ is either $\dot{D}_i^{3DMC}(x,y,z)$ or $\dot{D}_i^{DOM}(x,y,z)$, and $\sigma_i(x,y,z)$ is the combined statistical uncertainty of the entities in the numerator. The quantity $q_i$ is the dose difference at the i-th BNF point estimator spatial position between our "gold" standard - the BNF result, and the model result expressed in the units of the combined standard deviation.

If the differences were purely statistical, the probability distribution to find a point with a deviation $q$ would be given by the Gaussian distribution with mean of 0 and SD of 1.

$$p(q) = \frac{1}{\sqrt{2\pi}} \exp\left(-\frac{q^2}{2}\right). \tag{2}$$

However, if this distribution is not Gaussian, there are systematic differences between the two dose distributions. The systematic errors are separated from the statistical fluctuations by representing $q_i$ as the sum, $q_i = z_i + y_i$, of a systematic error, $z_i$ , and a statistical error, $y_i$ . We can treat $z_i$ and $y_i$ as random variables, distributed according to eq.(2) and *g(z)*, respectively. Since q is the sum of two random variables, its distribution, f(q), will be the convolution of p(z) and g(z)

$$f(q) = p(q) * g(q) = \int_{-\infty}^{\infty} ds'\, p(q-s')g(s'), \tag{3}$$

where '*' denotes convolution. The systematic error distribution, *g(z),* is assumed to have the following form

$$\begin{aligned} g(z) = {} & \alpha_1\left(a_1 z + b_1\right)\Theta(z-\Delta_1)\Theta(\Delta_2 - z) + \\ & \alpha_2\left(a_1 z + b_1\right)\Theta(z-\Delta_3)\Theta(\Delta_4 - z) + \left(1-\alpha_1-\alpha_2\right)\delta(z), \\ \Theta(z) = {} & \begin{cases} 0, & z<0 \\ 1, & z>0 \end{cases}. \end{aligned} \tag{4}$$

Since in the above expression only two out of the three parameters $\alpha_i$, $a_i$, $b_i$, ($i=1,2$) can be considered independent, the formula in eq.(4) has 8 independent parameters. It describes *g(z)* as a superposition of a fraction $\alpha_1$ of all points compared exhibiting systematic errors linearly distributed in the interval $[\Delta_1,\Delta_2]$, another fraction $\alpha_2$ showing systematic errors linearly distributed in the interval $[\Delta_3,\Delta_4]$, and a fraction $(1-\alpha_1-\alpha_2)$ which does not exhibit any systematic error. Eq.(4) is a generalization of the systematic distribution function used by Kawrakow and Fippel[32], who modeled the distribution of systematic errors as two Dirac delta functions. A graphical representation of *g(z)* is given in figure 3.

After normalizing eq.(4) to unity, and using and convolving it with the statistical error distribution, *f(q)* becomes

$$f(q)=\frac{\alpha_1}{\Delta_2-\Delta_1}\frac{1}{\sqrt{2\pi}}\exp\left(-\frac{q^2}{2}\right)\left\{a_1 I_1(\Delta_1,\Delta_2,q)+\left(1-\frac{a_1}{2}(\Delta_1+\Delta_2)\right)I_2(\Delta_1,\Delta_2,q)\right\}$$
$$+\frac{\alpha_2}{\Delta_4-\Delta_3}\frac{1}{\sqrt{2\pi}}\exp\left(-\frac{q^2}{2}\right)\left\{a_2 I_1(\Delta_3,\Delta_4,q)+\left(1-\frac{a_2}{2}(\Delta_3+\Delta_4)\right)I_2(\Delta_3,\Delta_4,q)\right\} \quad , \quad (5)$$
$$+\frac{(1-\alpha_1-\alpha_2)}{\sqrt{2\pi}}\exp\left(-\frac{q^2}{2}\right)$$

where

$$I_1(r,k,q)=\exp\left(-\frac{r^2}{2}+rq\right)-\exp\left(-\frac{k^2}{2}+kq\right)+q\cdot I_2(r,k,q),$$
$$I_2(r,k,q)=\exp\left(\frac{q^2}{2}\right)\sqrt{\frac{\pi}{2}}\left(erf\left[\frac{k-q}{\sqrt{2}}\right]-erf\left[\frac{r-q}{\sqrt{2}}\right]\right).$$

The error function, *erf, is* defined in the usual way as

$$erf(z)=\frac{2}{\sqrt{\pi}}\int_0^z \exp(-t^2)dt. \quad (6)$$

The function *f(q)* was fitted to the distribution eq.(1) using the Levenberg-Marquardt algorithm for least-squares fitting[33] as implemented in MINPACK[34]. Separate fits were performed at each of the six distances described in II.B.1. The resultant fits provided a model of systematic error for both 3DMC and DOM. We define the relative systematic error in percentage points *A%* at the position of the *i*-th BNF estimator as $A\%_i = 100\frac{A_{abs,i}}{\dot{D}_i}$ where $A_{abs,i}$ is the absolute systematic error in dose-rate units, and $\dot{D}_i$ is the dose-rate estimate. Then the fraction of all points at the given distance level with systematic errors between A% and B% is given by

$$F(A\%,B\%)=\frac{1}{N}\sum_{i=1}^{N}\int_{A\%/\sigma_i\%}^{B\%/\sigma_i\%} g(z)dz \quad (7)$$

where $N$=6335 is the number of BNF estimators in the level, and $\sigma_i\%$ is the relative percent standard deviation of 3DMC or DOM estimates at the position of the i-th BNF estimator. Figures 4 and 5 illustrate the applications of this deconvolution model to 3DMC and DOM data, respectively.

In addition to the convolution model, equation (5), the accuracy of 3DMC and DOM was quantified by direct comparison with MCBNF at each level through the mean of the overall root-mean square error

$$RMS(\%)=\sqrt{\frac{1}{N}\sum_{k=1}^{N}\left[1-\frac{\dot{D}_k^{Model}}{\dot{D}_k^{MCBNF}}\right]^2}\cdot 100, \quad (8)$$

where again $N$=6335 is the number of BNF estimators in each level. Eq.(8) provides a quantitative measurement for both systematic and stochastic errors combined.

We employed both eqs.(7) and (8) to compare efficiency of the DOM and 3DMC. We required that the two methods satisfy the following accuracy criteria. Since distances beyond 2 cm are of less clinical significance, we relaxed accuracy requirements in this region.

1. **For distances < 2 cm:** *F(-3%,3%)>0.90,RMS(%)≤3%* leading to accuracy of 5% or better in about 90% of the points analyzed as it will be seen in section III. Note that *RMS(%)* constraints apply to the total (stochastic + systematic) error.
2. **For distances ≥ 2 cm:** *F(-10%,10%)>0.90, RMS(%)≤5%,* leading to accuracy of better than 10% in more than 95% of the points analyzed.

In the present study we refer to the *F(A,B)* condition as the systematic error criterion while the *RMS(%)* condition is referred to as the overall root mean-square error criterion.

All 8 parameters of the systematic error distribution function eq.(5) are expressed in units of the standard deviation resulting from the statistical error of the simulation, as seen from eq.(1). Since the stochastic and systematic errors are independent from each other, smaller statistical uncertainties would support estimation of the *g(z)* parameters with better accuracy. Consequently, the estimate of the integral systematic error eq.(7) would be of better accuracy as well. Thus the high statistics DOM and 3DMC simulations, which have statistical uncertainties of the order of 0.5%, were used for the estimation of the systematic error distribution, g(z), parameters. Because the high statistics g(z) is independent of the number of simulated histories, it describes the systematic error characteristic of the low statistics simulations. Thus low statistics simulations were used only to define the minimum number of histories needed to satisfy the overall root mean-square overall error criterion78.

**II.E Determination of Simulation Run-Time Parameters**

To fairly compare DOM and 3DMC CPU times leading to relative efficiency estimates, it necessary to identify DOM and MC simulation parameters that yield equivalent accuracy performance.

In the case of 3DMC, two variables were adjusted: the voxel sizes at each of the 6 distances and the number of histories. For the fixed distance of each BNF detector set, *k*, we calculated the average efficiency $\varepsilon$ of the voxel grid applied by

$$\varepsilon_k = \frac{1}{s_k^2 T} \text{ where } s_k = \frac{1}{M_k}\sum_{i=1}^{M_k}\frac{\sigma_{k,i}^{3DMC}}{\dot{D}_{k,i}^{3DMC}}, \tag{9}$$

where $T$ is the 3DMC CPU time in hours, $M_k$ is the number of BNF estimators in the level, $\dot{D}_{k,i}^{3DMC}$ is the 3DMC interpolated value at the *i*-th BNF point-estimator location in the level, *k*. and $\sigma_{k,i}^{3DMC}$ is the corresponding standard error about the mean. For each level, we sought to identify the largest voxel sizes (hence the highest efficiency) that satisfies the systematic error criterion, e.g., *F(-3%,3%)>0.90* for distance < 2 cm, in subsection II.D. To this end, the original basic 0.1 and 1 mm voxels were merged into larger voxels to achieve better statistics when accuracy was not compromised. An automated procedure was developed which was merging the voxels from the basic grid according to

$$\dot{D}_E^{3DMC} = \frac{1}{\sum_{i,j,k}\Delta V_{ijk}}\sum_{i,j,k}\Delta V_{ijk}\dot{D}_{ijk}^{3DMC},$$

$$\sigma_E^{3DMC} = \frac{1}{\sum_{i,j,k}\Delta V_{ijk}}\sqrt{\sum_{i,j,k}\Delta V_{ijkijk}^2\left(\sigma_{ijk}^{3DMC}\right)^2} \tag{10}$$

where $\dot{D}_E^{3DMC}$ is the dose rate estimate for the merged voxel and $\sigma_E^{3DMC}$ is its corresponding standard deviation about the mean. The quantities $\dot{D}_{ijk}^{3DMC}$ and $\sigma_{ijk}^{3DMC}$ are the scored dose rate and its standard error about the mean in dose-rate units, respectively, from the voxel *i, j, k,* of the original grid, where $\Delta V_{ijk}$ is the corresponding voxel volume. The summation is over the neighboring voxels being merged. It will be seen in section III that this is an important step in reducing the CPU time and improving the efficiency of our 3DMC simulations. For each of the 6 six distance groups, the high statistics simulation was used to determine the maximal extent of voxel merging consistent with the systematic error criteria of section II.D.

Once the 3DMC merged voxel grids were established, the number of histories in each 'low statistics' simulation, as discussed in subsection II.A, was determined by finding smallest number of histories that satisfies the root-mean square error criterion described in subsection II.D. The requisite number of histories was found to be 2,650,000 and 330,000 for 0.1 mm and 1 mm basic voxel grids, respectively. The results are discussed in detail in section III.

Similar to the 3DMC parameter optimization, the high statistics (1,613,400 first collision-source rays, yielding total kerma statistical precisions of less than 1%) DOM run was used to define systematic error distribution. By trial-and-error, we refined the finite-differencing spatial mesh until the systematic error criterion (subsection II.D) was achieved at each BNF level. This resulted in 40x34x35 and 32x32x33 spatial mesh grids for the large and small distances simulations, respectively.

Then, the number of first-collision source rays[30] was varied to find the minimum number needed to satisfy the overall root-mean-square error criterion (subsection II.D) along with the associated CPU time required. This process yielded 214,320 rays for the 40x34x35 grid (distances greater than or equal to 1 cm) and 63,600 for the 32x32x33 grid (distances less than 1 cm) in order to accommodate the overall root-mean-square error criterion. These DOM simulations are referred to below as low statistics simulations.

## *III. RESULTS AND DISCUSSION*

All calculations below were carried out on a SGI workstation with 4 RS12000 195MHz processors. As noted in section II, all simulations use a single processor, and all CPU timing results are relative to this workstation. Naturally, using faster processors would lead to smaller CPU times for each simulation regardless of the method used.

### III.A. 3DMC and DOM systematic error analysis with the implementation of high statistics.

For the high statistics 3DMC simulations, Figures 4a and 4b compares the histogram of measured relative error histogram to the convolution model for two different distances (corresponding to two different BNF shells of 6335 points each). These graphs show the systematic distribution *g(z)* and demonstrate that equation (5) accurately describes the total error distribution of for voxel detectors at distances of 0.25 cm and 1 cm. Figures 5a and 5b plot the same data in the case of DOM.

Tables 2 and 3 show the results of the error analysis for all distances studied. Table 2 shows the 3DMC results for all voxel sizes studied. It is evident that voxel grid size of 0.1 mm satisfies the systematic error criterion at all distances starting at 0.25 cm. Increasing the voxel size to 0.2 mm violates the systematic error criterion at 0.25 cm but is completely suitable for larger distances. Similarly, 1 mm voxel size is sufficient for adequate results at distances of 1 cm or larger. Voxel sizes of 2 mm or 3 mm, while violating the systematic error criterion at 1 cm distance, meet the criterion far from the seed. Table 3 shows that the spatial mesh discretization used by both the short- and long-distance PARTISN simulations satisfy the systematic error requirements at all distances. Figures 4 and 5 show that DOM exhibits smaller fraction of data entries with negligible systematic error (the last term in eq.(4)) compared to 3DMC – 0.041 vs 0.629 at 0.25 cm and 0.193 vs 0.529 at 1 cm distance, respectively. This in turn leads to smaller in magnitude systematic error estimate for $F(-3\%,+3\%)$ and $F(-10\%,+10\%)$ in the case of DOM relative to 3DMC. A major cause for this occurrence is the variable size of the spatial mesh cells used in the DOM simulations, which results in larger variation in both the statistical and systematic errors at a given distance (level) for DOM relative to 3DMC.

Table 2 demonstrates the complex behavior of the error distribution with distance and voxel size for our 3DMC simulations. For instance, it is seen that while 0.1 mm voxel grid is adequate in terms of our systematic error criterion, the number of histories analyzed for high statistics (9,000,000) is insufficient to sustain the overall root mean-square criterion even at 0.5 cm and therefore more histories are needed which consequently would lead to larger CPU times. However, due to the adopted procedure for merging of the basic voxel grids eq.(1) it is evident that 0.2 mm grid satisfies the root mean-square criterion and therefore no more histories are needed. Thus the merging procedure effectively reduces the overall CPU time and leads to better 3DMC efficiency in our analysis as seen from the comparisons of the efficiency $\varepsilon$ of each simulation also included in Table 2. We do not consider this a violation of our goal for objective 3DMC/DOM efficiency comparison since we expect that the current restriction for uniform grid sizes in our 3DMC calculations will be overcome in the near future without significant reduction in the efficiency provided by the use of the Siddon algorithm[23].

### III.B. 3DMC and DOM simulations, overall root mean-square error criterion implementation for low statistics.

The results of our 3DMC runs presented in table 2 show that when the systematic error criterion is satisfied for a given voxel grid size, the violation of the overall root mean-square error criterion can occur only if the stochastic error becomes sufficiently large, which may occur at larger distances. Such larger stochastic errors, however, can be reduced sufficiently by using larger voxel sizes at larger distances without violating the systematic error criterion. Thus, 0.1 mm voxels satisfy both the total and systematic terror criteria at 0.25 cm distance while 1 mm voxels are adequate at 1 cm distance, and can be used to establish the overall CPU time needed for 3DMC calculations. However, the same is not true for our DOM simulations. Due to variable size of the spatial mesh grid, the overall root mean-square error analysis should be carried out for each distance in the same manner as for the systematic error in the previous section.

For DOM and 3DMC, we progressively reduced the number of first collision rays or total histories, respectively After each reduction, the RMS(%) error was evaluated at each distance (for DOM) or at the shortest distance (for 3DMC) until the overall root mean-square criterion was satisfied. In this way we arrived at the minimum “low statistics” number of stochastic iterations needed for both DOM and 3DMC to meet the RMS error criterion at all distances simultaneously.

Table 4 shows the resulting RMS(%) at each distance for the two 3DMC low statistics simulations. 2,650,800 histories are necessary to meet the RMS(%) error criterion of 3.0% at the 0.25 and 0.5 cm distances using voxel sizes of 0.1 to 0.2 mm. For larger distances, 330,000 histories are required for voxel sizes ranging from 1 to 4 mm. This, in conjunction with the systematic error criterion fulfillment from Table 2 shows that 3DMC achieves the required accuracy with the two basic voxel grids after a total of 2,980,800 histories with CPU=1.65 hours.

Table 5 shows the same results but for the two grids used in our DOM simulations. Unlike the 3DMC, the number of first collision source rays had to be chosen such that the RMS(%) magnitude to satisfy the root mean-square error criterion simultaneously at each distance. Table 5 shows that the root mean-square error criterion has been satisfied for the 40x34x35 grid at all three distances analyzed. It was necessary to exceed the criterion at 0.25 cm and 0.50 cm distances in order to satisfy it at 1.0 cm. Similar results for the 32x32x33 grid and the remaining distances were achieved. Thus, in conjunction with systematic error criterion fulfillment from Table 3 shows that DOM achieves the required accuracy with the two grids with the use of 277,920 rays for the first collision source and with total CPU=0.74 hours.

To illustrate the accuracy achieved by each method the tables include the cumulative fraction of all BNF/Model comparisons for which the quantity $\left|1-\left(\dot{D}^{Model}/\dot{D}^{MCBNF}\right)\right|\cdot 100$ is larger than 5% and 10%. The tables show that the implementation of the accuracy criteria leads to accuracy of better than 5% in about 90% of the BNF point estimator locations for distances smaller than 2 cm, and better than 10% in 95% of all BNF point estimator locations for distances of 2 cm or larger.

Figures 6 through 8 plots the fraction of calculation points exceeding the specified absolute percent error [eq.(11)] at distances of 0.25 cm, 1.0 cm, and 2.0 cm for both DOM and 3DMC. At 0.25 cm (Figure 6), the frequency of DOM errors is smaller than that of 3DMC until the 5% discrepancy level is reached. This is because DOM must outperform the equivalent accuracy criterion at 0.25 cm to satisfy them at larger distances. The difference is within a few percent and both DOM and 3DMC methods demonstrate similar accuracy for larger discrepancies beyond the 5% point. Figures 7 and 8 demonstrate that 3DMC and DOM performance is much more closely matched at distances 1 cm and 2 cm, respectively. Figure 8 illustrates the importance of voxel merging on 3DMC-accuracy performance. Given the DOM and 3DMC CPU times given in Tables 4 and 5, it is evident that in this setting, DOM is about 2-fold more CPU-efficient than 3DMC.

This gain in efficiency is significantly smaller than our previous results[5] where DOM simulations were 73-fold more CPU-efficient relative to analog scoring EGS4 calculations of the same dimensionality. The major reason for this improved 3DMC performance is the higher efficiency of expected-value tracklength scoring compared to analog estimation. Hedtjarn et al[12] have demonstrated that use of tracklength scoring improves Monte Carlo efficiency 20- to 100-fold when a 3D grid of uniform size voxel detectors is used.

The error model developed in the present study supports improved optimization of computational efficiency by decoupling selection of voxel size and spatial discretization schemes needed to minimize systematic errors and selection of optimal number of histories or first collision source rays needed to minimize stochastic errors. The results indicate that:

- In contrast to our previous 2D DOM simulations, where the first-collision source dominated algorithm efficiency, the first collision source ray-tracing algorithm takes only about 30% of the overall DOM PARTISN CPU time;
- For 3DMC simulation, the use of voxel sizes that vary with distance is critical for optimizing efficiency.

Future developments in both DOM and 3DMC simulation algorithms could change the two-fold efficiency gain of DOM relative to Monte Carlo reported here. For example, variance reduction techniques such as correlated sampling[12], use of Sobol number sequences[32], and more efficient ray tracing[36] have been shown to increase Monte Carlo efficiency by at least a factor of two. On the other hand, implementation of recently developed algorithms such as the adaptive mesh refinement[25] (AMR), and second and higher-order spatial/angular differencing schemes[26], are expected to significantly improve DOM efficiency. Also, the extent to which DOM and MC codes can be parallelized is an important issue.

Regardless of the potential of each method for future improvements, it is clear that DOM codes have the potential to support computational efficiencies in three-dimensional geometries that are comparable to or better than the most efficient low-energy $^{125}$I-brachytherapy Monte Carlo photon-transport solutions reported to date. Future work will address whether this improved or comparable better efficiency can be sustained for higher energy brachytherapy sources such as $^{192}$Ir or $^{137}$Cs and whether adaptive mesh refinement and other DOM improvements can further increase computational efficiency. In addition, the error model proposed here maybe used as a foundation to an automated process of optimization of the spatial/energy/angular discretization utilizing the actual patient anatomy.

[Paragraph added 07/04/26]. While the ca. 2003 computer hardware, specific code versions, and referenced accuracy employed by this study are outdated, a major original contribution of the paper is to outline an approach for defining equivalent performance between deterministic and stochastic algorithms that compute scalar quantities distributed over a 3D grid. The core of the approach is to model the distribution of errors, relative to a low-uncertainty bench simulation, achieved through the grid, allowing systematic and statistical errors to be separated. Then separate criteria for satisfactory levels of systematic and random error, applicable to both Monte Carlo, deterministic and hybrid algorithms can be identified, enabling “fair” assessment of computation efficiency. In addition, the paper demonstrates simulation parameters can be separately optimized: by manipulating non-uniform grids for the former and number histories/stochastic rays for the latter. The method as presented requires a high-statistics benchmark with negligible systematic error, BNFMC in the case of this study. A fruitful direction of future study would be assessment of the generality of the run-time and grid optimization results within a class of problems.

## IV. CONCLUSIONS

The present study compares the efficiency of three-dimensional deterministic discrete ordinates PARTISN calculations with voxel-based Monte Carlo simulations of the dose distribution around a model 6702 $^{125}$I seed in a homogeneous water phantom in three-dimensional Cartesian geometry. A procedure for defining equivalent accuracy, based upon separation of stochastic and systematic errors has been developed.

Based upon our analysis, the PARTISN discrete ordinates code is a factor of two more CPU efficient relative to the most efficient voxel-based Monte Carlo code available to us at least. in the case of low energy $^{125}$I brachytherapy. Our empirical results are consistent with the theoretical analysis of Börgers[35], who concludes that deterministic algorithms employing finite-difference schemes with order of accuracy higher than 1 (which is the case for the adaptive-weighted diamond differencing scheme used in this study) compete well with Monte Carlo.

## *V.ACKNOWLEDGMENTS*

The authors would like to express their appreciation and gratitude to Dr. I. Kawrakow from the NRC of Canada for the valuable discussion and recommendations at different stages of this work. The present study was supported by the National Institutes of Health grant R01 CA46640.

Figure Captions.

**Figure 1.** (a): The geometry representation of the $^{125}I$ 6702 seed by the Monte Carlo: Material composition: ion-exchange resin ($C_{12}H_{18}NCl$) spheres (density $\rho$=1.2 g·cm$^{-3}$) with uniformly distributed $^{125}I$ throughout their volume; Ti capsule ($\rho$=4.54 g·cm$^{-3}$); air with composition by weight - 75.53% N, 23.18% O, 1.29% Ar, with density of $\rho$=1.20x10$^{-3}$ g·cm$^{-3}$. All sizes in mm. (b): BNF point estimators arrangement in a level – all level estimators at the same distance r from the seed center – both $\theta$ and $\Psi$ vary from 0º to 180º.

**Figure 2.** Three-dimensional spatial discretization of the $^{125}I$ 6702 seed used in our PARTISN DOM simulations. All sizes in cm.

**Figure 3.** Graphical representation of the general form of the systematic density function eq.(6).

**Figure 4.** Systematic error deconvolution using the model eqs. (5) - (7) for the 3DMC simulations at two distances with two voxel sizes: **(a)**- error distribution at 0.25 cm distance for simulation with a voxel size of 0.2 mm, average relative SD of 0.1%; **(b)**- same as in **(a)** for a simulation at 1 cm with a voxel size of 2 mm, average relative SD of 0.2%. Dots – raw data; solid curves – fitted function $f(q)$ eq.(7); dashed curves – the resulting $g(z)$ eq.(6). 3DMC high statistics.

**Figure 5.** Same as in figure 4 for the case of DOM: **(a)**- at 0.25 cm distance, average relative SD of about 0.1%; **(b)**- at 1 cm distance – average relative SD of about 0.3%. DOM high statistics.

**Figure 6.** Cumulative fraction of all BNF points compared at 0.25cm for the DOM and 3DMC. The DOM calculation using a 40x34x35 grid and low statistics (214,320 rays). The 3DMC simulations using 0.2 mm voxel dimensions and low statistics (2,650,800 histories). The ordinate of each point of the histograms shows the cumulative fraction of all BNF/Model comparisons with relative difference larger than or equal to the abscissa of the same point.

**Figure 7.** Same as in figure 6 for distance of 1.0 cm. DOM results for the grid 40x34x35 with low statistics (214,320 rays). 3DMC with 1 mm voxel grid size, low statistics (330,000 histories).

**Figure 8.** Same as in figure 7 for distance of 2.0 cm. DOM results for the grid 32x32x33 with low statistics (63,600 rays). 3DMC 1 mm denotes simulations with 1 mm voxel grid size, low statistics (330,000 histories), 3DMC 2 mm denotes result of the same 3DMC simulation after merging the 1 mm voxels into 2 mm voxels according to eq.(1).

**Table 1.** Standard deviation (SD) of the mean of the primary and total kerma calculated by PARTISN using the stochastic ray-tracing first collision source. PARTISN is run $N$ = 51 times with the only difference being a different random seed number for the ray-tracing algorithm. The SD is calculated as[31] $SD_A = \sqrt{\frac{1}{N(N-1)}\left[\sum_{i=1}^{N} x_i - \frac{1}{N}\sum_{i=1}^{N}(x_i)^2\right]}$ while the relative SD as $SD_R = \sqrt{\frac{N}{N-1}\left[\frac{\sum_{i=1}^{N} x_i}{\left(\sum_{i=1}^{N} x_i\right)^2} - \frac{1}{N}\right]}$. Data present for different primary-to-total kerma ratios at various distances.

| Distance (cm) | Primary/Total kerma ratio | Primary kerma | | Total kerma | |
|---|---|---|---|---|---|
| | | $SD_A$ (x$10^{-3}$) | Relative $SD_R$ (x$10^{-2}$) | $SD_A$ (x$10^{-3}$) | Relative $SD_R$ (x$10^{-2}$) |
| 0.25 | 0.879 | 25.4 | 0.1396 | 25.5 | 0.1233 |
| 1.00 | 0.460 | 2.85 | 1.053 | 2.87 | 0.4883 |
| 2.00 | 0.321 | 0.794 | 1.790 | 0.798 | 0.5784 |
| 3.00 | 0.104 | 0.194 | 4.323 | 0.196 | 0.4550 |

**Table 2.** 3DMC simulation efficiency with high statistics and error analysis for different voxel sizes. Voxels of 1 mm size or larger are used for distances of 1 cm or larger while smaller voxel sizes are used for the other distances. *F(±A%)* denotes *F(-A%,A%)* (eq.(9)) and *RMS(%)* is given by eq.(10).

| distance (cm) | Voxel Size | | | | | | | | | | | | | | | | | |
|---|---|---|---|---|---|---|---|---|---|---|---|---|---|---|---|---|---|---|
| | 0.1 mm | | | 0.2 mm | | | 0.3 mm | | | 1 mm | | | 2 mm | | | 3 mm | | |
| | *F(±3%)* | *RMS(%)* | $\varepsilon$ | *F(±3%)* | *RMS(%)* | $\varepsilon$ | *F(±3%)* | *RMS(%)* | $\varepsilon$ | *F(±10%)* | *RMS(%)* | $\varepsilon$ | *F(±10%)* | *RMS(%)* | $\varepsilon$ | *F(±10%)* | *RMS(%)* | $\varepsilon$ |
| 0.25 | 1.00 | 2.06 | 152 | 0.89 | 2.06 | 1222 | 0.86 | 3.50 | 3915 | - | - | - | - | - | - | - | - | - |
| 0.50 | 1.00 | 3.13 | 39 | 0.99 | 1.59 | 330 | 0.98 | 1.26 | 1122 | - | - | - | - | - | - | - | - | - |
| 1.00 | - | - | - | - | - | - | - | - | - | 1.00* | 1.32 | 856 | 0.85* | 3.03 | 7205 | 0.85* | 5.70 | 21826 |
| 2.00 | - | - | - | - | - | - | - | - | - | 1.00 | 1.85 | 268 | 1.00 | 1.20 | 2214 | 1.00 | 1.38 | 7716 |
| 3.00 | - | - | - | - | - | - | - | - | - | 1.00 | 2.63 | 127 | 1.00 | 1.33 | 1084 | 1.00 | 1.04 | 3484 |
| 5.00 | - | - | - | - | - | - | - | - | - | 1.00 | 4.42 | 40 | 1.00 | 2.28 | 333 | 1.00 | 1.58 | 1123 |

*The values calculated are those of *F(-3%,3%)* for the distance of 1 cm.

**Table 3.** DOM error analysis with high statistics for both spatial grids. The systematic error calculation *F*(*±A%*)=*F*(*-A%,A%*) and the overall root mean-square error *RMS(%)* are given for both spatial grid meshes used for distances closer (40x34x35) and larger (32x32x33) than 1 cm.

| Distance /cm | 40x34x35 grid | | 32x32x35 grid | |
|---|---|---|---|---|
| | *F(±3%)* | *RMS(%)* | *F(±10%)* | *RMS(%)* |
| 0.25 | 0.90 | 1.55 | - | - |
| 0.50 | 0.90 | 1.79 | - | - |
| 1.00 | 0.90 | 1.87 | - | - |
| 2.00 | - | - | 0.97 | 2.11 |
| 3.00 | - | - | 0.96 | 2.72 |
| 5.00 | - | - | 0.95 | 3.51 |

**Table 4.** 3DMC analysis for low statistics. The CPU time/number of histories for each grid are listed only once since the result is valid for all distances spanned by it. Rows marked ">5%" or ">10%" denote the cumulative fraction of all BNF/3DMC comparisons for which the quantity $\left|1-\frac{\dot{D}^{3DMC}}{\dot{D}^{MCBNF}}\right|\cdot 100$ is larger than 5% or 10%, respectively (see figure 6). Results are valid for SGI RS12000 195 MHz single processor workstation.

| | **0.1 mm grid** | | **1.0 mm grid** | | | | **Total** |
|---|---|---|---|---|---|---|---|
| | **0.25 cm 0.1 mm voxel** | **0.50[a)] cm 0.2 mm voxel** | **1.00 cm 1 mm voxel** | **2.00[b)] cm 2 mm voxel** | **3.00[b)] cm** | **5.00[c)] cm** | |
| **RMS(%)** | 3.006 | 2.95 | 2.971 | 2.798 | 3.880 | 3.884 | - |
| **>5%** | 0.090 | 0.091 | 0.0847 | 0.106 | 0.232 | 0.199 | |
| **>10%** | $3.192\text{x}10^{-3}$ | $1.756\text{x}10^{-3}$ | $3.033\text{x}10^{-3}$ | 0.000 | 0.015 | 0.015 | |
| **# histories** | 2,650,800 | - | 330,000 | - | - | - | 2,980,800 |
| **CPU (hours)** | **1.15** | **-** | **0.50** | **-** | **-** | **-** | **1.65** |

**Table 5.** DOM results for low statistics. # rays denotes the number of rays analyzed during the run of the first collision source for each grid. Rows marked “>5%” or “>10%” denote the cumulative fraction of all BNF/3DMC comparisons for which the quantity $\left|1-\frac{\dot{D}^{DOM}}{\dot{D}^{MCBNF}}\right|\cdot 100$ is larger than 5% or 10%, respectively (see figure 7). The numbers in brackets in the CPU row are the CPU (in hours) for the first collision source calculation. Results are valid for SGI RS12000 195 MHz single processor workstation.

| | **40x34x35 grid** | | | **32x32x33 grid** | | | **Total** |
|---|---|---|---|---|---|---|---|
| | **0.25 cm** | **0.50 cm** | **1.00cm** | **2.00 cm** | **3.00 cm** | **5.00 cm** | |
| **RMS(%)** | 2.542 | 2.296 | 3.06 | 3.444 | 3.674 | 4.481 | - |
| **>5%** | 0.0803 | 0.112 | 0.110 | 0.096 | 0.141 | 0.287 | - |
| **>10%** | $5.401\text{x}10^{-3}$4 | $7.544\text{x}10^{-3}$ | 0.012 | 0.017 | 0.016 | 0.0327 | - |
| **# rays** | 214,320 | - | - | 63,600 | - | - | 277,920 |
| **CPU/h** | **0.46 (0.14)** | **-** | **-** | **0.28 (0.08)** | **-** | **-** | **0.74 (0.21)** |

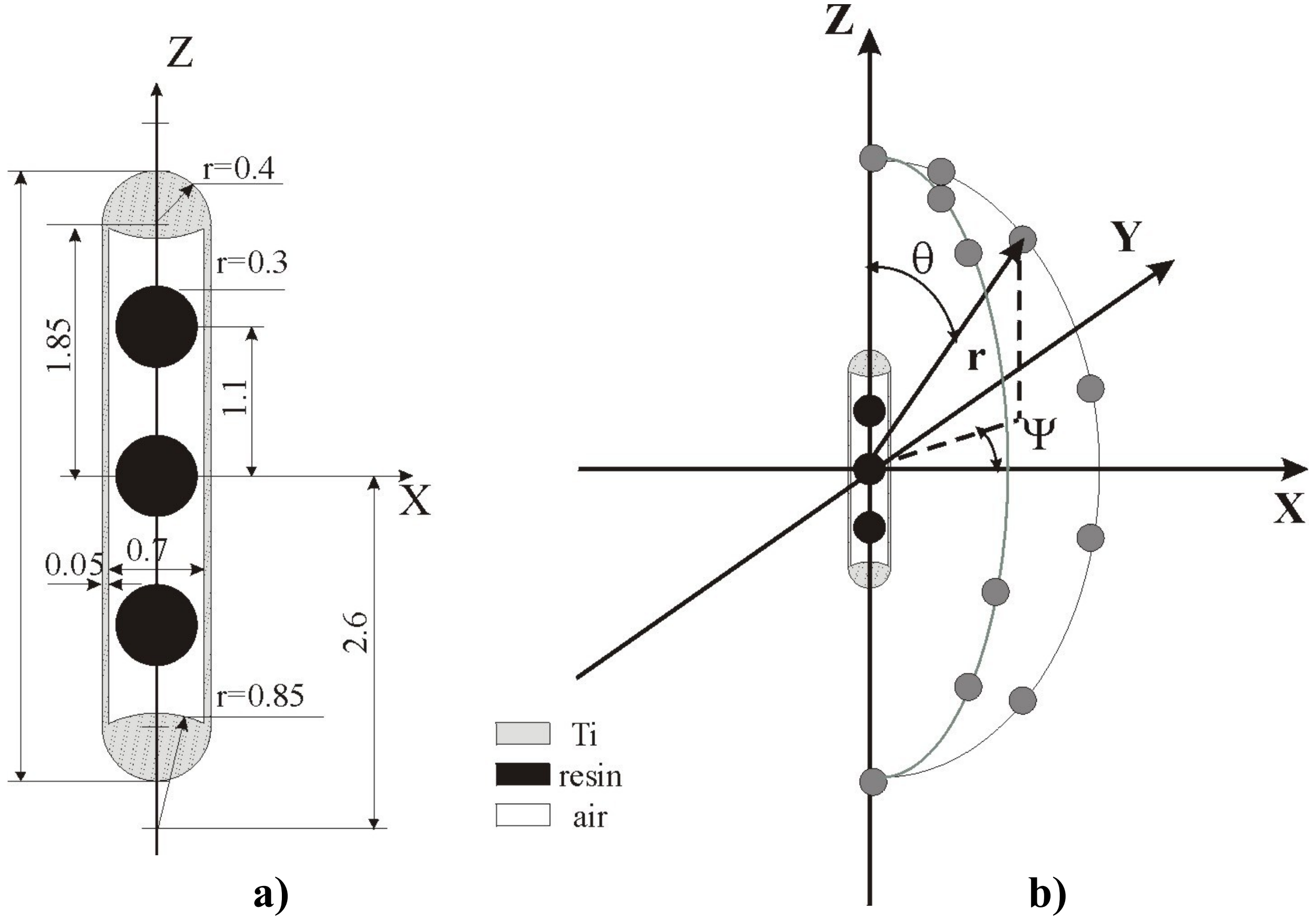

**Figure 1.** (a): The geometry representation of the $^{125}$I 6702 seed by the Monte Carlo: Material composition: ion-exchange resin ($C_{12}H_{18}NCl$) spheres (density $\rho$=1.2 g·cm$^{-3}$) with uniformly distributed $^{125}$I throughout their volume; Ti capsule ($\rho$=4.54 g·cm$^{-3}$); air with composition by weight - 75.53% N, 23.18% O, 1.29% Ar, with density of $\rho$=1.20x10$^{-3}$ g·cm$^{-3}$. All sizes in mm. (b): BNF point estimators arrangement in a level – all level estimators at the same distance r from the seed center – both θ and Ψ vary from 0º to 180º.

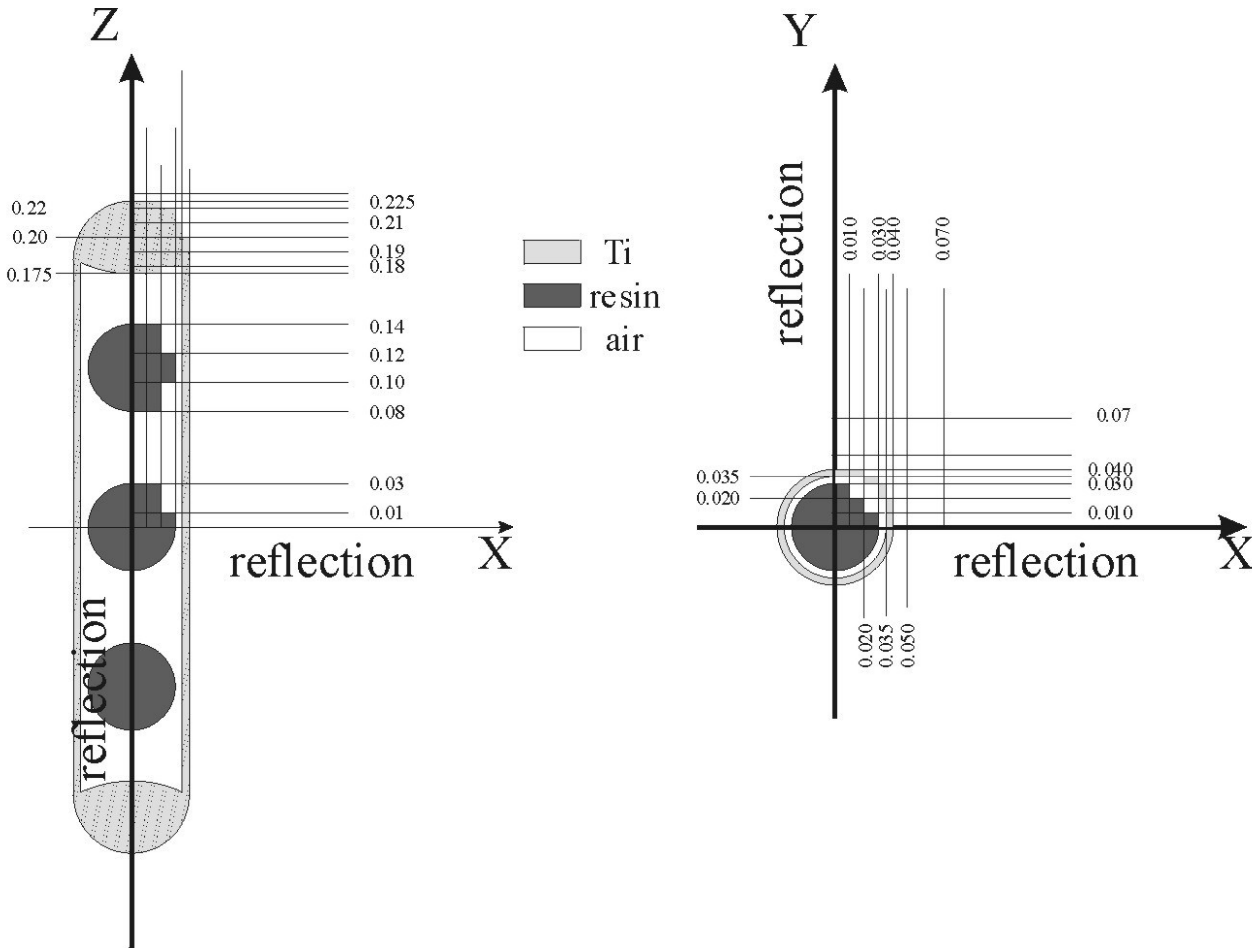

**Figure 2.** Three-dimensional spatial discretization of the $^{125}$I 6702 seed used in our PARTISN DOM simulations. All sizes in cm.

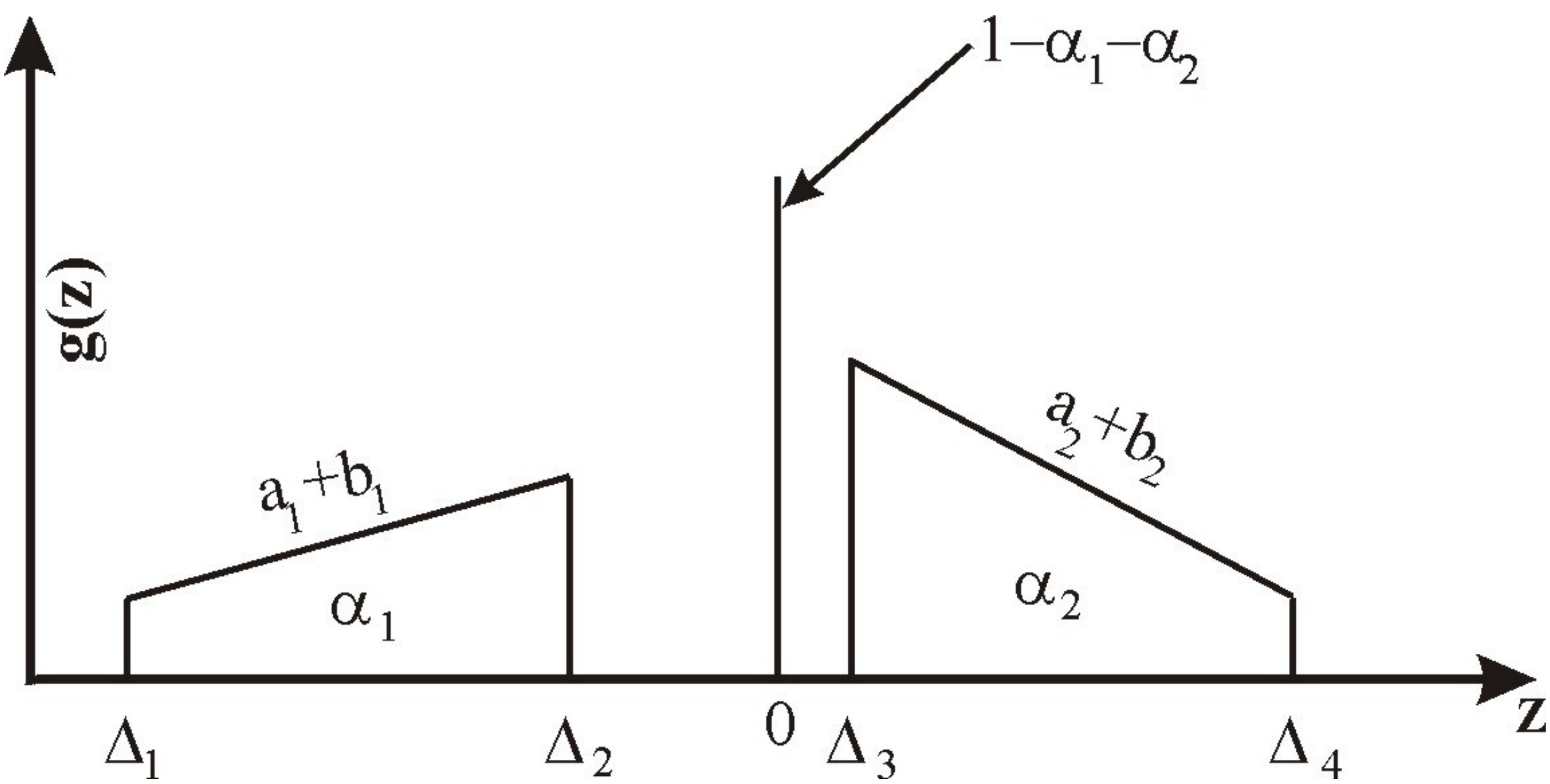

**Figure 3.** Graphical representation of the 8-parameter normalized to unity systematic density function g(z) eq.(4).

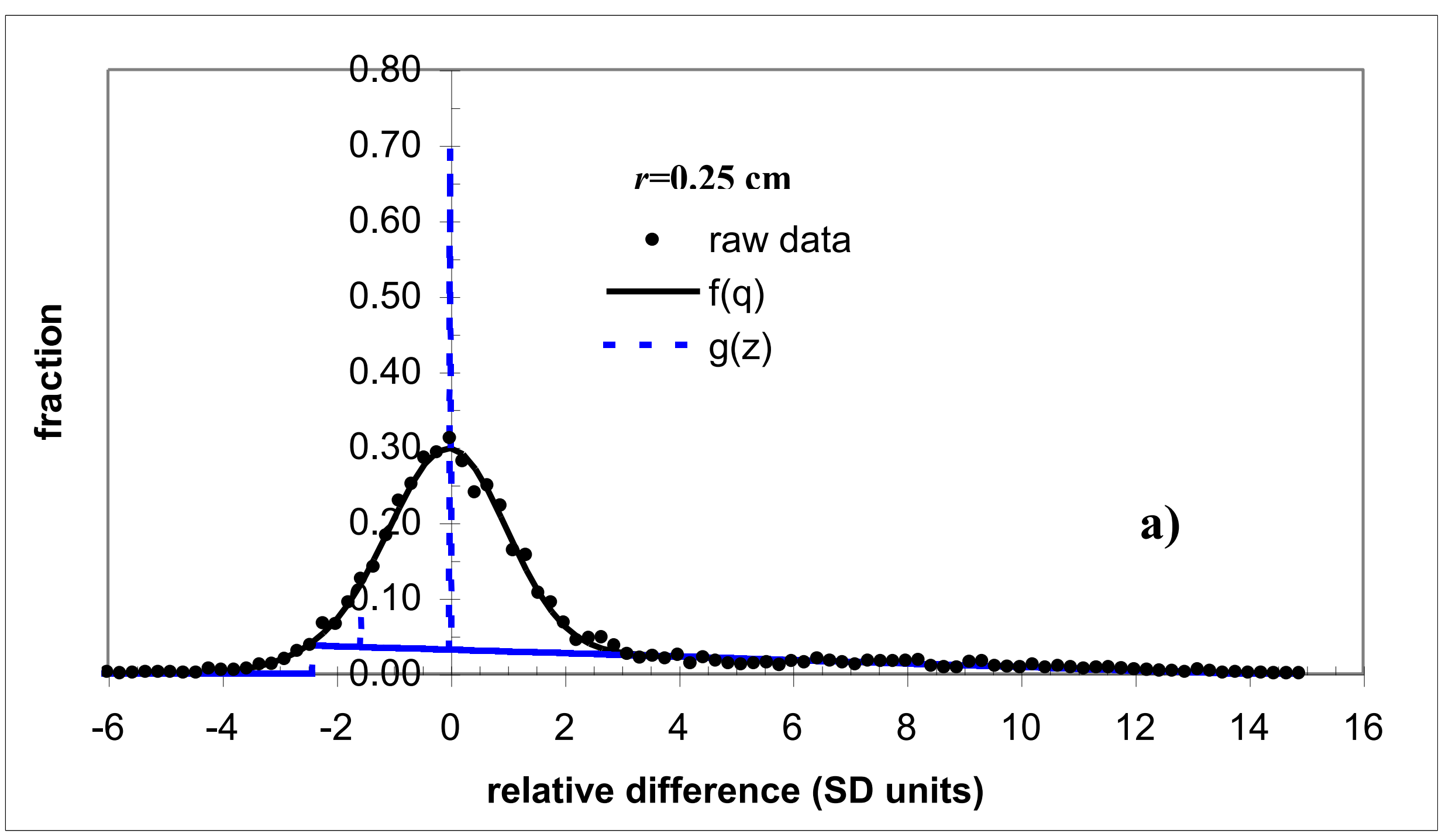

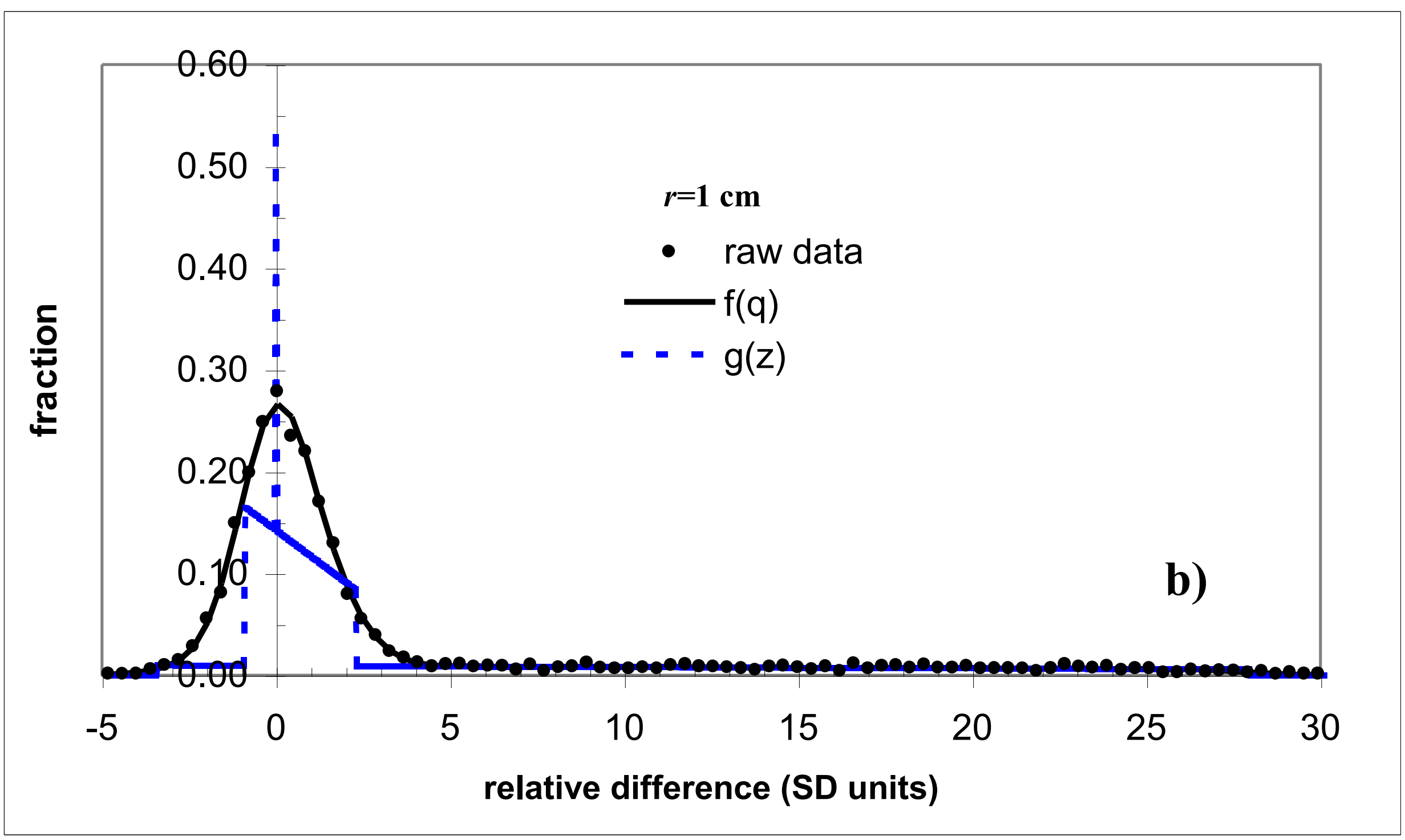

**Figure 4.** Systematic error deconvolution using the model eqs. (5) for the 3DMC simulations at two distances with two voxel sizes: **(a)**- error distribution at 0.25 cm distance for simulation with a voxel size of 0.2 mm, average relative SD of 0.1%; **(b)**- same as in **(a)** for a simulation at 1 cm with a voxel size of 2 mm, average relative SD of 0.2%. Dots – raw data; solid curves – fitted function *f(q)* eq.(5); dashed curves – the resulting *g(z)* eq.(4). 3DMC high statistics.

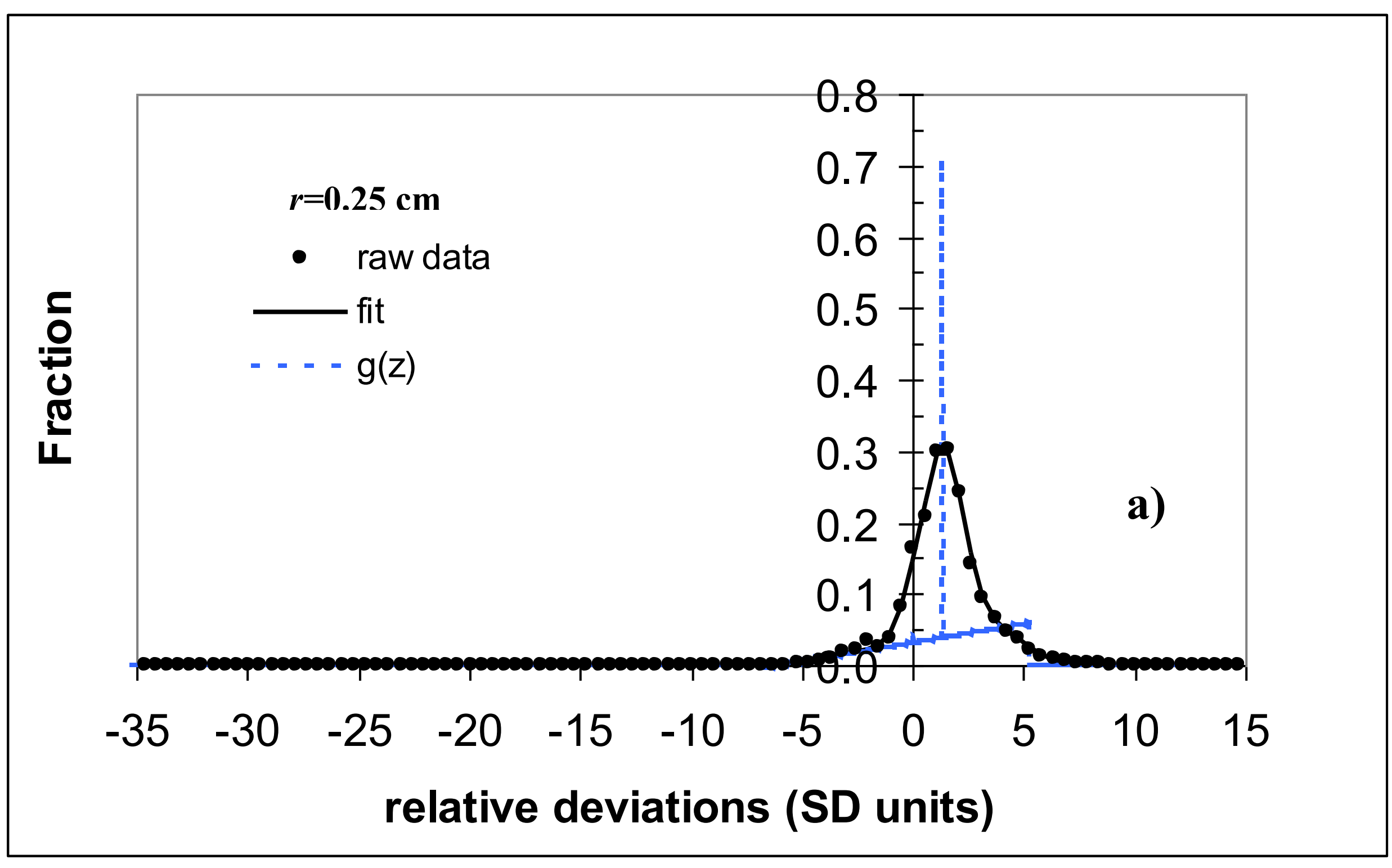

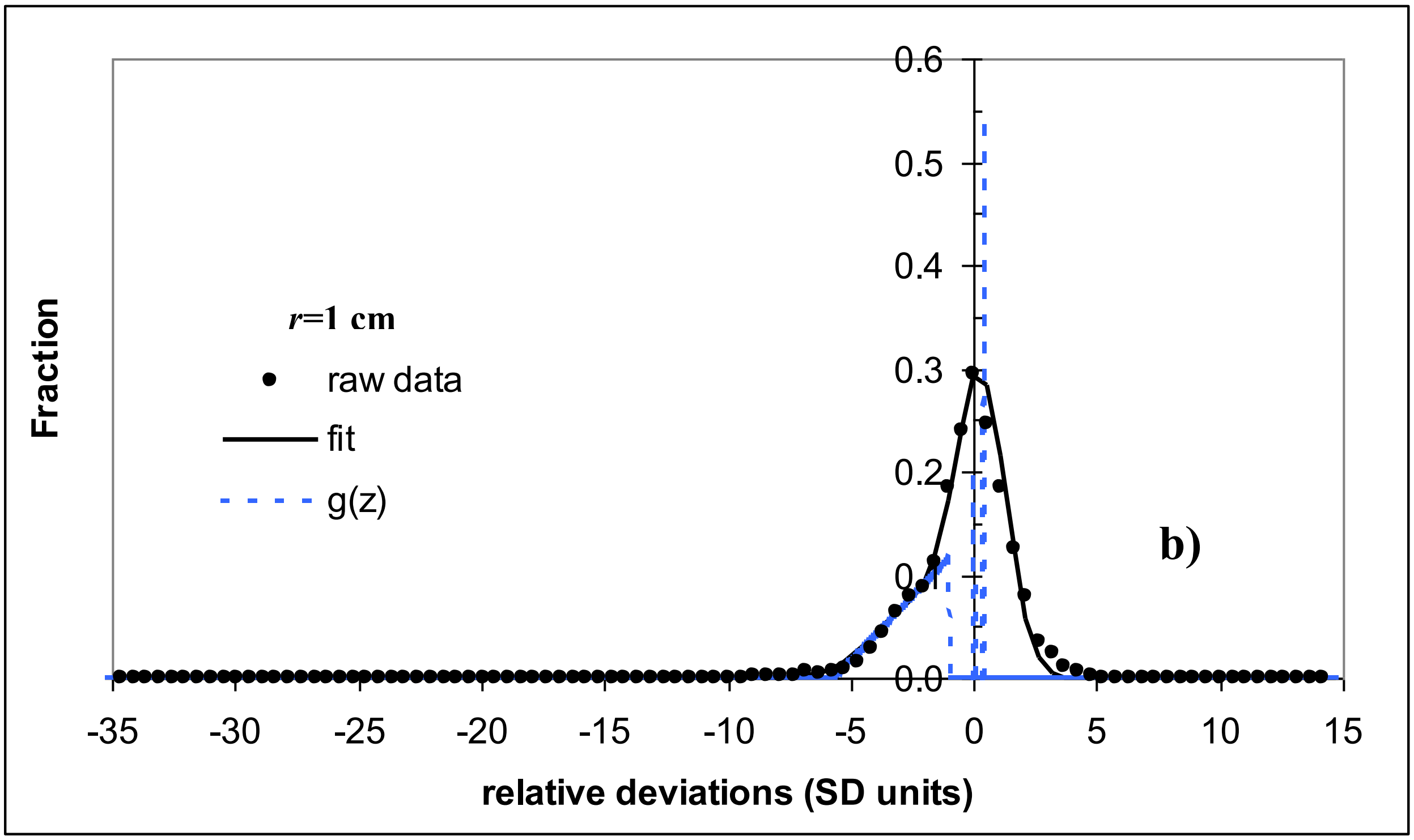

**Figure 5.** Same as in figure 4 for the case of DOM: **(a)**- at 0.25 cm distance, average relative SD of about 0.1%; **(b)**- at 1 cm distance – average relative SD of about 0.3%. DOM high statistics.

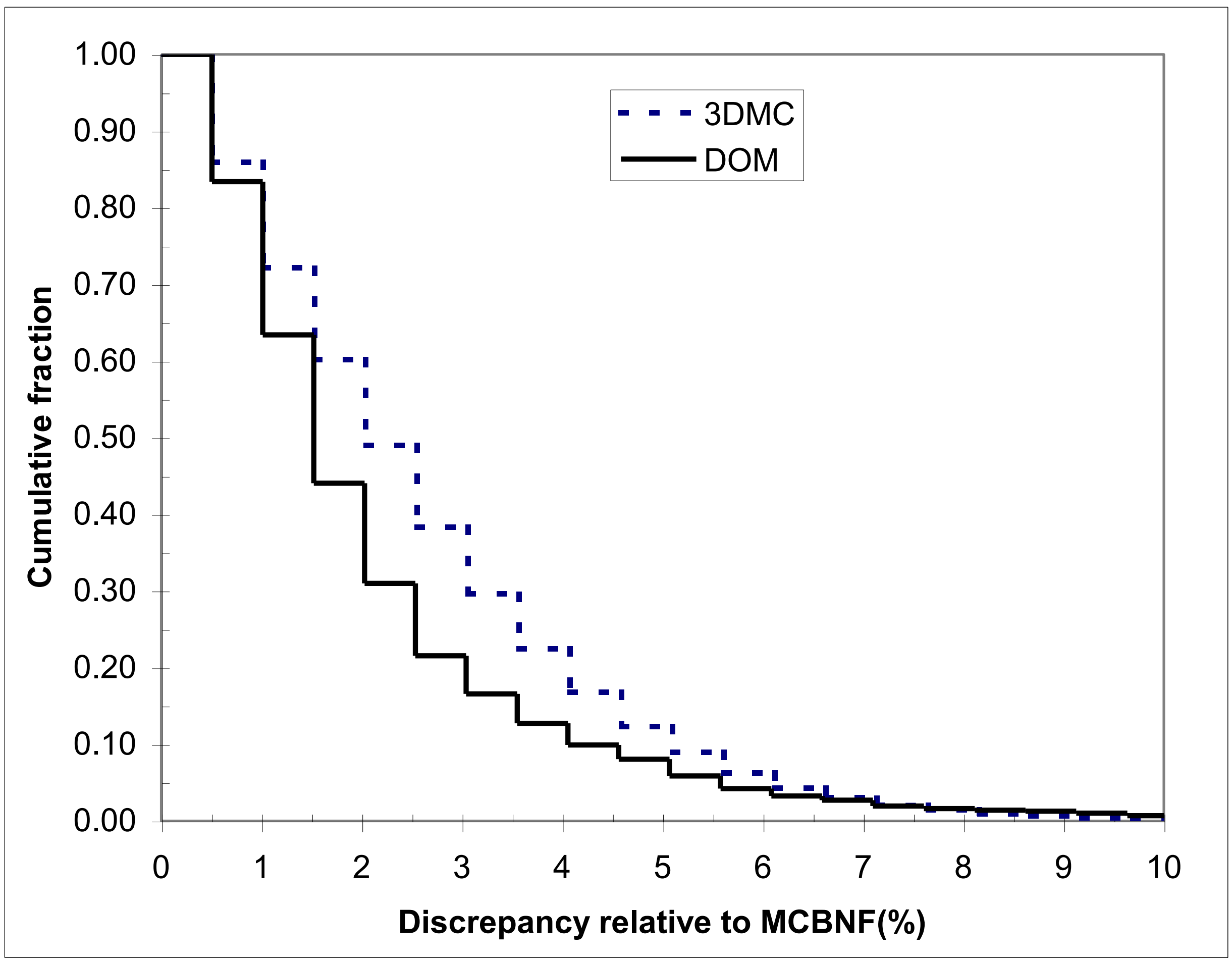

**Figure 6.** Cumulative fraction of all BNF points at 0.25cm for DOM and 3DMC that deviates from MCBNF by more than the indicated percentage on the abscissa. The DOM calculation used the 40x34x35 grid and low statistics (214,320 rays). The 3DMC simulations used 0.2 mm voxel dimensions and low statistics (2,650,800 histories). The ordinate of each point of the histograms shows the cumulative fraction of all BNF/Model comparisons with relative difference larger than or equal to the abscissa of the same point.

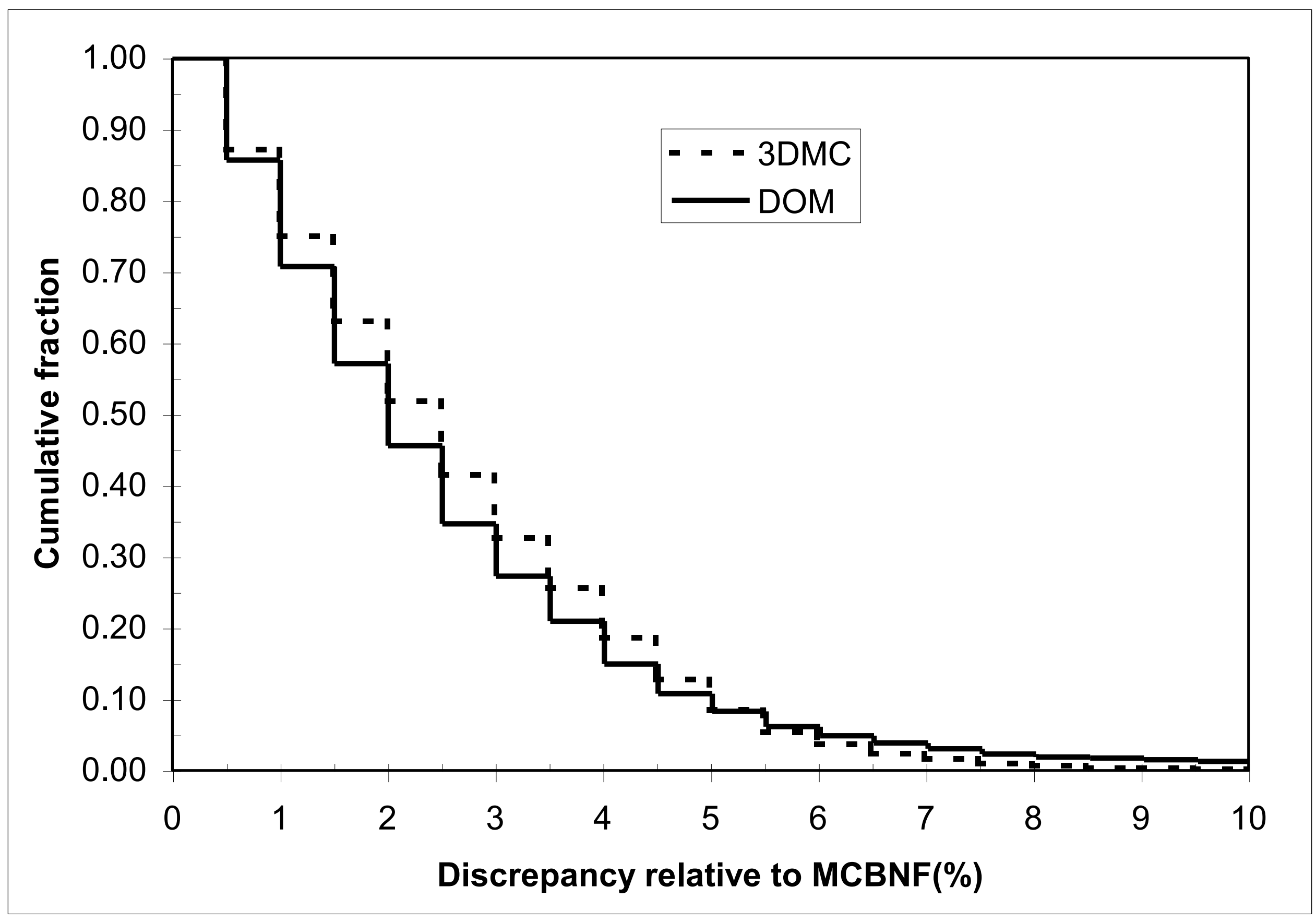

**Figure 7.** Same as in figure 6 for distance of 1.0 cm. DOM results for the 40x34x35 grid with low statistics (214,320 rays). 3DMC with 1 mm voxel grid size, low statistics (330,000 histories).

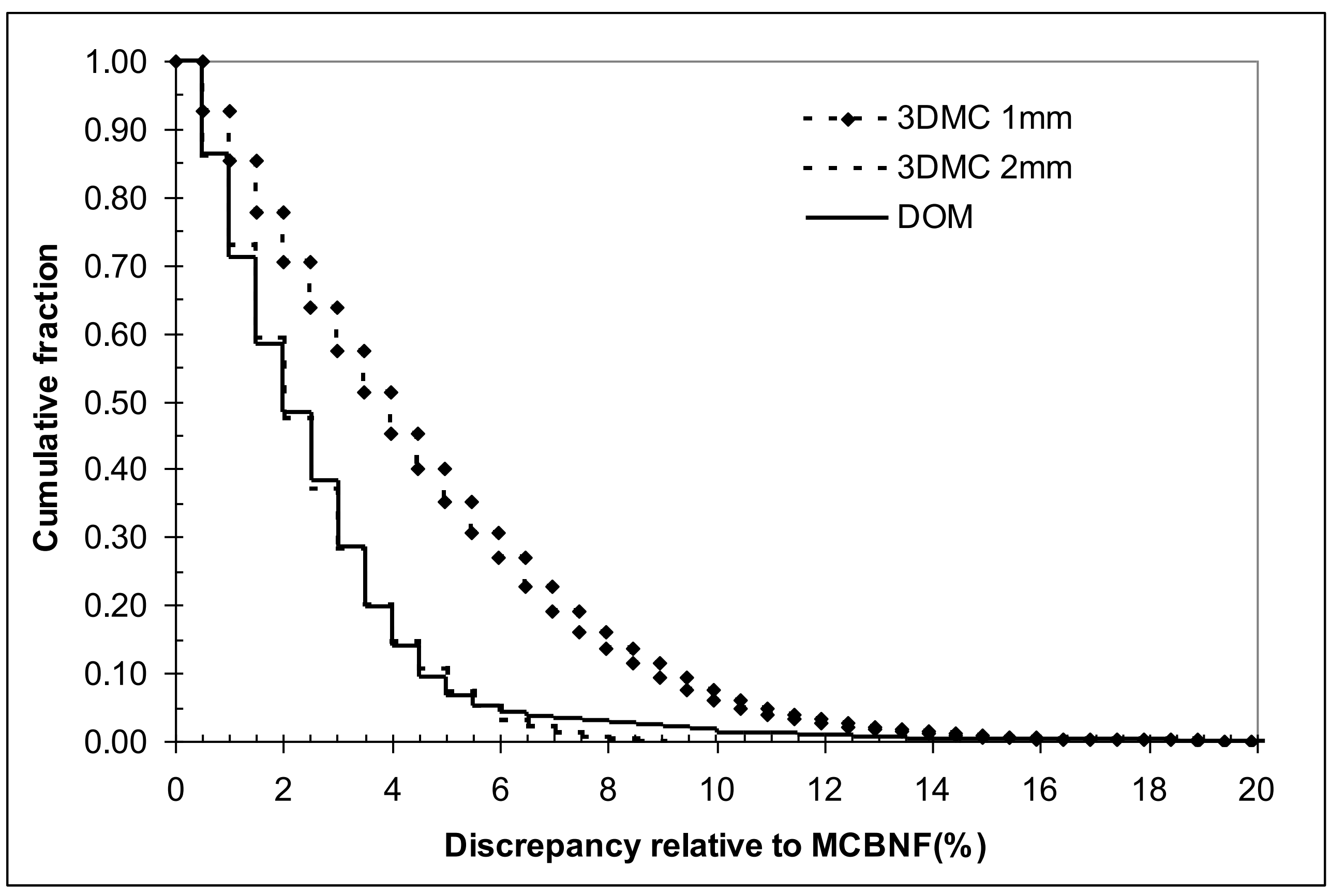

**Figure 8.** Same as in figure 7 for distance of 2.0 cm. DOM results for the 32x32x33 grid with low statistics (63,600 rays). 3DMC 1 mm denotes simulations with 1 mm voxel grid size, low statistics (330,000 histories), 3DMC 2 mm denotes result of the same 3DMC simulation after merging the 1 mm voxels into 2 mm voxels according to eq.(10).